\documentclass[12pt]{article}
\usepackage{graphicx,amssymb,amsfonts}
\newlength{\extraspace}
\newlength{\extraspaces}
\newcounter{savefootnote}
\begingroup
\endgroup

\newcommand{\be}{\begin{equation}
\addtolength{\abovedisplayskip}{\extraspaces}
\addtolength{\belowdisplayskip}{\extraspaces}
\addtolength{\abovedisplayshortskip}{\extraspace}
\addtolength{\belowdisplayshortskip}{\extraspace}}
\newcommand{\ee}{\end{equation}}

\newcommand{\ba}{\begin{eqnarray}
\addtolength{\abovedisplayskip}{\extraspaces}
\addtolength{\belowdisplayskip}{\extraspaces}
\addtolength{\abovedisplayshortskip}{\extraspace}
\addtolength{\belowdisplayshortskip}{\extraspace}}
\newcommand{\ea}{\end{eqnarray}}

\newcommand{\newsection}[1]{
\vspace{7mm} \pagebreak[3] \addtocounter{section}{1}
\setcounter{subsection}{0} 
\begin{center}
{\large {\bf \thesection. #1}}
\end{center}
\nopagebreak
\medskip
\nopagebreak \hspace{3mm}}

\begin{document}

\begin{center}
{\bf    Regularization of $f(T)$ gravity theories and local Lorentz transformation}\footnote{ \hspace*{.5cm}
 }
\end{center}
\begin{center}
{\bf Gamal G.L. Nashed}\footnote{ Mathematics Department, Faculty of Science, Ain
Shams University, Cairo, 11566, Egypt \\
\hspace*{.2cm} Egyptian Relativity Group (ERG) URL:
http://www.erg.eg.net}
\end{center}

\bigskip

\centerline{\it Centre  for Theoretical Physics, The British
University in  Egypt} \centerline{\it Sherouk City 11837, P.O. Box
43, Egypt }

We regularized the field equations of $f(T)$ gravity theories such that
 the effect of Local Lorentz Transformation (LLT), in the case of spherical symmetry, is removed. A  ``general
tetrad field'', with  an arbitrary function of radial coordinate preserving
  spherical symmetry is provided. We split that tetrad field
into two matrices; the first represents a LLT,
which contains an arbitrary function, the second matrix
represents a proper tetrad field which  is a solution to the field
equations of $f(T)$ gravitational theory, (which are not invariant under
 LLT).  This  ``general tetrad field'' is then applied
to the regularized field equations of $f(T)$. We show that the effect
of the arbitrary function which is involved in the LLT  invariably disappears.

\newsection{\bf Introduction}

Amended gravitational  theories  have become very interesting  due to their
ability to provide an alternative framework for understanding the nature of dark
energy. This is done through the modifications of the gravitational Lagrangian so as it render  an arbitrary function of its original argument, for instance
$f(R)$ instead of Ricci scalar  $R$ in the Einstein-Hilbert action \cite{NO1,NO2,NO3,DM}.

Indeed there exists an equivalent construction of General Relativity (GR) dependent
on the concept of parallelism. The idea is initially done by  Einstein who had tried to make  a unification
between electromagnetism and gravity fields using absolute
parallelism spacetime \cite{Ea,Ea1}. This goal was frustrated by the lack of  a Schwarzschild solution. Much later, the theory
of  absolute parallelism gained much attention as a modification  theory of gravity, refereed to ``teleparallel equivalent of general relativity" (TEGR) (cf. \cite{Mc,Mc1,Mc2,Mc3,Mc4}).
  The basic block in TEGR is the tetrad field.
  The tetrad field  consists of fields of orthonormal bases which belong
to the tangent space of the manifold.
Note that the contravariant tetrad field, ${h_i}^\mu$, has sixteen components while the metric tensor has only ten. However,
the tetrads are invariant under local Lorentz rotations.

The aim of the modification is to treat a more general manifold
which comprises in addition to curvature a quantity called ``torsion''.
The curvature tensor, consists of a part without torsion
plus part with torsion, is  vanishing identically. One
can generally use either the torsion-free part or the torsion
part to represent the gravitational field. The most suitable
 way is to deal with the covariant tetrad field, ${h^i}_\mu$, and the so-called
Weitzenb\"ock spacetime \cite{Wr}. The tetrad field describes fields of orthonormal bases, which are related to the tangent spacetime
 of the manifold with spacetime coordinates $x^\mu$. This tangent spacetime is Minkowski spacetime  with the
metric $\eta_{i j}$ that can be defined at any given point on the
manifold.

Recently, modifications   of TEGR have been studied  in the
domain of cosmology \cite{FF7,FF8,FF9}. This is  known as $f(T)$
gravity and is built from a generalized Lagrangian \cite{FF7,FF8,FF9}.
In such a theory, the gravitational field is not characterized  by curved
spacetime but with torsion. Moreover, the field equations are only
second order unlike the fourth order equations of the $f(R)$ theory.

Many of $f(T)$ gravity theories had been analyzed in
\cite{BF9}-\cite{WY1}. It is found that $f(T)$ gravity theory is not dynamically
equivalent  to TEGR  Lagrangian
through  conformal transformation \cite{Yr1}. Many observational
constraints  had been studied  \cite{Bg1}-\cite{WMQ}. Large-scale
structure in $f(T)$ gravity theory had been analyzed \cite{LSB,LSB1};
perturbations in the area of cosmology in $f(T)$ gravity had been
demonstrated \cite{SNH}-\cite{CCDDS}; Birkhoff$'$s theorem,
 in $f(T)$ gravity had  been studied \cite{MW1}. Stationary solutions having spherical symmetry have
been derived for  $f(T )$ theories \cite{Wt1,Ngrg,Nprd,Nas}. Relativistic stars
and the cosmic expansion  derived in \cite{Dy,SSFR}.

 Nevertheless,  a major problem of $f(T)$ gravitational theories is that they are not locally Lorentz invariant and appear to
harbour extra degrees of freedom.

The goal of  this study is to regularize the field equations
of $f(T)$ gravitational theory so that we remove the effect of
LLT.  We then apply   a ``general tetrad'' field,  Which is  consists of two matrices the first is a solution to the non-invariant field equation of $f(T)$ and the second matrix is a local Lorentz transformation, to the amended  field equations and show that the effect
of LLT disappears.

 In \S 2, a brief survey of the $f(T)$ gravitational theory is presented.

In \S 3, a ``general tetrad'' field, having spherically symmetric  with an arbitrary
 function of the radial coordinate $r$ is applied  to the  field  equations  of
 the  $f(T)$   which are not invariant under LLT.
  It is shown that the arbitrary function has affect in this application.

    In \S, 4, we derive the field equations  of $f(T)$  which are invariant under LLT.
    We then apply these amended field equations to a ``general tetrad'' field. We show  the effect of the arbitrary function   invariably disappears.

    The final section is devoted to discussion.

\newsection{Brief review of f(T)}

In the Weitzenb\"ock spacetime, the fundamental field variables
describing gravity are a quadruplet of parallel vector fields
\cite{Wr} ${h_i}^\mu$, which we call the tetrad field. This is
characterized by: \begin{equation} D_{\nu} {h_i}^\mu=\partial_{\nu}
{h_i}^\mu+{\Gamma^\mu}_{\lambda \nu} {h_i}^\lambda=0, \end{equation} where
${\Gamma^\mu}_{\lambda \nu}$ defines the nonsymmetric affine
connection:  \begin{equation} {\Gamma^\lambda}_{\mu \nu} \stackrel{\rm def.}{=}
{h_i}^\lambda {h^i}_{\mu, \nu}, \end{equation} with $h_{i \mu, \
\nu}=\partial_\nu h_{i \mu}$\footnote{spacetime indices $\mu, \ \
\nu, \cdots$ and SO(3,1) indices a, b $\cdots$ run from 0 to 3. Time
and space indices are indicated by $\mu=0, i$, and $a=(0), (i)$.}.

Equation (1) leads to the metricity  condition and the identical
vanishment  of the curvature tensor defined by
${\Gamma^\lambda}_{\mu \nu}$, given by equation (2). The metric
tensor $g_{\mu \nu}$
 is defined by
 \begin{equation} g_{\mu \nu} \stackrel{\rm def.}{=}  \eta_{i j} {h^i}_\mu {h^j}_\nu, \end{equation}
with $\eta_{i j}=(-1,+1,+1,+1)$ that is metric of Minkowski
spacetime. We note that, associated with any tetrad field
${h_i}^\mu$ there  is a metric field defined
 uniquely by (3), while a given metric $g^{\mu \nu}$ does not
 determine the tetrad field completely,  any LLT of the tetrad  ${h_i}^\mu$ leads to a new set of
 tetrad  which also satisfies (3).

The torsion components and the contortion are defined as: \begin{eqnarray}
{T^\alpha}_{\mu \nu}  & \stackrel {\rm def.}{=} &
{\Gamma^\alpha}_{\nu \mu}-{\Gamma^\alpha}_{\mu \nu} ={h_a}^\alpha
\left(\partial_\mu{h^a}_\nu-\partial_\nu{h^a}_\mu\right),\nonumber\\
{K^{\mu \nu}}_\alpha  &\stackrel {\rm def.}{=} &
-\frac{1}{2}\left({T^{\mu \nu}}_\alpha-{T^{\nu
\mu}}_\alpha-{T_\alpha}^{\mu \nu}\right), \end{eqnarray}    where the contortion
equals the difference between Weitzenb\"ock  and Levi-Civita
connection, i.e., ${K^{\mu}}_{\nu \rho}= {\Gamma^\mu}_{\nu \rho
}-\left \{_{\nu  \rho}^\mu\right\}$.

 One can defined the skew-symmetric  tensor ${S_\alpha}^{\mu \nu}$  as \begin{equation} {S_\alpha}^{\mu \nu}
\stackrel {\rm def.}{=} \frac{1}{2}\left({K^{\mu
\nu}}_\alpha+\delta^\mu_\alpha{T^{\beta
\nu}}_\beta-\delta^\nu_\alpha{T^{\beta \mu}}_\beta\right),\end{equation} which
is skew symmetric in the last two indices. The torsion scalar is
defined as \begin{equation} T \stackrel {\rm def.}{=} {T^\alpha}_{\mu \nu}
{S_\alpha}^{\mu \nu}. \end{equation}Similar to the $f(R)$ theory, one can
define the action of $f(T )$ theory as \begin{eqnarray}   {\cal
L}({h^a}_\mu)=\int d^4x h\left[\frac{1}{16\pi}f(T)\right], \qquad \qquad \textrm {where} \quad
h=\sqrt{-g}=det\left({h^i}_\mu\right),\end{eqnarray}  (assuming   units
in which $G = c = 1$).   Considering the action in equation (7) as a
function of the fields ${h^i}_\mu$ and putting  the variation of the
function with respect to the field ${h^i}_\mu$ to be vanishing, one
can obtain the following equations of motion \cite{CGSV,BF9}: \begin{equation}
{S_\mu}^{\rho \nu} T_{,\rho} \
f(T)_{TT}+\left[h^{-1}{h^i}_\mu\partial_\rho\left(h{h_i}^\alpha
{S_\alpha}^{\rho \nu}\right)-{T^\alpha}_{\lambda \mu}{S_\alpha}^{\nu
\lambda}\right]f(T)_T-\frac{1}{4}{\delta^\nu}_\mu f(T)=-4\pi {{\cal
T}^\nu}_{\mu},\end{equation}
 where
$T_{,\rho}=\frac{\partial T}{\partial x^\rho}$,
$f(T)_T=\frac{\partial f(T)}{\partial T}$,
$f(T)_{TT}=\frac{\partial^2 f(T)}{\partial T^2}$ and ${{\cal
T}^\nu}_{\mu}$ is the
energy momentum tensor.

 In this study we are
interested in studying the vacuum case of $f(T)$ gravity theory,
i.e., ${{\cal
T}^\nu}_{\mu}=0$.
\newsection{Spherically symmetric solution in  f(T) gravity theory}

Assuming that the manifold is a stationary and spherically
symmetric $\left( {h^i}_\mu \right)$ has the form:
\begin{equation}
\left({h^i}_\mu \right) = \left( \matrix{ LA+HA_2 &LA_1+HA_3 &0 & 0
\vspace{3mm} \cr -(LA_2+HA) \sin\theta \cos\phi & -(LA_3+HA_1)
\sin\theta \cos\phi & - r\cos\theta \cos\phi & r\sin\theta\sin\phi
\vspace{3mm} \cr -(LA_2+HA) \sin\theta \sin\phi & -(LA_3+HA_1)
\sin\theta \sin\phi & - r\cos\theta \sin\phi & -r\sin\theta\cos\phi
\vspace{3mm} \cr -(LA_2+HA) \cos\theta & -(LA_3+HA_1) \cos\theta &
r\sin\theta
 &  0 \cr}\right), \end{equation}
where $A(r)$, $A_1(r)$, $A_2(r)$ and  $A_3(r)$ are four unknown functions of the radial coordinate  $r$,
 $L=L(r)=\sqrt{H(r)^2+1}$ and $H=H(r)$ is an arbitrary function.
Tetrad fields (9) transform as \[ \left({h^i}_\mu \right)=\left({\Lambda^i}_j\right) \left({h^j}_\mu\right)_1,\]
where $\left({h^j}_\mu\right)_1$ is given by
\begin{equation} \left( {h^j}_\mu \right)_1= \left( \matrix{A(r) & A_1(r)& 0 &
0\vspace{3mm} \cr A_2(r)\sin\theta \cos\phi &A_3(r)\sin\theta
\cos\phi&r\cos\theta \cos\phi & -r\sin\theta \sin\phi \vspace{3mm}
\cr A_2(r)\sin\theta \sin\phi&A_3(r)\sin\theta \sin\phi&r\cos\theta
\sin\phi & r\sin\theta \cos\phi \vspace{3mm} \cr A_2(r)\cos\theta &
A_3(r)\cos\theta&-r\sin\theta & 0 \cr } \right).\end{equation}
 The tetrad field (10) has been studied \cite{Ngrg} and it has been shown that the  solution to the $f(T)$ gravitational
 theory  has the form:
\begin{equation} A=1-\displaystyle\frac{M}{r}, \qquad A_1=\displaystyle\frac{M}{r(1-\displaystyle\frac{M}{r})},
\qquad A_2=\displaystyle\frac{M}{r}, \qquad
A_3=\displaystyle\frac{1-\displaystyle\frac{M}{r}}{1-\frac{2M}{r}},\end{equation} where $M$ is
the gravitational mass.  Equation   (11) is an exact vacuum solution to field equations
of $f(T)$ gravitational theory  provided that \begin{equation} f(0)=0,
 \qquad f_T(0)\neq0, \qquad f_{TT}\neq0.\end{equation}

The LLT $\left({\Lambda^i}_j\right)$ has the form:
\begin{eqnarray} &&\left({\Lambda^i}_j\right) =   \left( \matrix{
L & H \sin\theta \cos\phi & H \sin\theta \sin\phi &  H \cos\theta
\vspace{3mm} \cr - H \sin\theta \cos\phi & 1+H_1\sin\theta^2 \cos\phi^2 &H_1\sin\theta^2
\sin\phi \cos\phi &H_1\sin\theta \cos\theta \cos\phi
\vspace{3mm} \cr - H \sin\theta \sin\phi &H_1
\sin\theta^2 \sin\phi \cos\phi &1+H_1\sin\theta^2
\sin\phi^2 &H_1\sin\theta \cos\theta \sin\phi
\vspace{3mm} \cr - H \cos\theta &H_1\sin\theta
\cos\theta \cos\phi &H_1\sin\theta \cos\theta \sin\phi
&1+H_1\cos\theta^2 \cr}\right).\nonumber\\
&&
\textrm {where} H_1=\left(L-1 \right). \end{eqnarray}
From the general spherically symmetric local Lorentz transformation, Eq.~ (13), one can generate the previous spheriaclly symmetric solution \cite{SNH}.

    Using Eq.~(11)  in Eq.~(9),
   one can obtain $h = \det ({h^\mu}_a) =r^2\sin\theta$
   and, with the use of Eqs. (4) and (5), we obtain the
torsion scalar and its derivatives in terms of r \begin{eqnarray} &&
T(r)=\frac{4([1-MH']L+HH'[M-r]-L^2)}{r^2L},
\qquad \textrm{where} \qquad H'=\frac{\partial H(r)}{\partial r},
\nonumber\\
&& \nonumber\\
&&  T'(r)=\frac{\partial T(r)}{\partial
r}=-\frac{4\{rL^2H''[(r-M)H+ML]-r(M-r)H'^2-2MH'L^2[L-H]+2L^3(1-L)\}}{r^3L^3}.\nonumber\\
&&\end{eqnarray}  The field
equations (7)  have the form \begin{eqnarray} && 4\pi{{\cal
T}_0}^0=-\frac{f_{TT}T'[M(2+H)+L(r-M)-r]}{r^2}
+\frac{f_{T}}{r^2L}\left[L(1-MH')+HH'(M-r)-L^2\right]+\frac{f}{4}\,\nonumber\\
&&  \end{eqnarray}
\begin{equation} 4\pi{{{\cal
T}_1}}^0=\frac{4f_{TT}T'[(M-r)H-ML]}{r(r-2M)},\end{equation}
\begin{equation} 4\pi{{\cal T}_1}^1=\frac{f_{T}
\{(1-MH')L+HH'(M-r)-L^2
\}}{r^2L}+\frac{f}{4}\,\end{equation}
\begin{eqnarray} && 4\pi{{\cal T}_2}^2=4\pi{{\cal
T}_3}^3=-\frac{f_{TT}T'\{M(1+H)-r+L(r-M)\}}{2r^2}
\nonumber\\
&&+\frac{f_{T}\{(1-MH')L+(M-r)HH'-L^2\}}{r^2L}+\frac{f}{4}. \end{eqnarray}
Equations (14)-(18), show that the field  equations of $f(T)$ are effected by the inertia
which is located in the LLT given by Eq.~(13). This effect  is related to
the non-invariance of the field equations of $f(T)$ gravitational theory under LLT.
\newsection{Regularization of f(T) gravitational theory under LLT}
The tetrad field $\left({h^i}_{  \mu} \right)$ transforms under LLT as:
\begin{equation} \left({{{ \bar  h}}^i}_{ \ \mu} \right)=\left({\Lambda^i}_j(x)\right) \left({h^j}_{  \mu} \right).\end{equation}
The derivatives of $\left({{ \bar h}^i}_{\  \mu} \right)$ has the form:
\begin{equation} \frac{\partial \left({{ \bar h}^i}_{ \ \mu} \right)}{\partial x^\nu}=\left({{ \bar h}^i}_{\ \mu,\; \nu} \right)=
\left({\Lambda^i}_j(x)\right)_{, \; \nu}\left({h^j}_{  \mu} \right)+\left({\Lambda^i}_j(x)\right)\left({h^j}_{\mu,\;\nu} \right)
.\end{equation} The non-symmetric affine connection
constructed from the tetrad field $\left({{{ \bar  h}}^i}_{ \ \mu} \right)$ has the form
\begin{equation} {{\bar \Gamma}^\mu}_{\ \nu \rho
}= \eta^{i j}\left({{{ \bar  h}}_i}^{ \ \mu} \right)\left({{ \bar h}}_{\ j \nu,\;\rho} \right).\end{equation}
Using equations (19) and (20) in (21) one gets
\begin{equation} {{\bar \Gamma}^\mu}_{\ \nu \rho}= {{\Gamma}^\mu}_{\nu \rho}+\left({\Lambda_j}^i(x)\right)\left({h_i}^{ \ \mu} \right)\left({\Lambda^j}_k(x)\right)_{,\;\rho}\left({h^k}_{\nu} \right),\end{equation}
where ${{\Gamma}^\mu}_{\nu \rho}$ is the non-symmetric affine connection constructed from
the tetrad field  $\left({h_i}^{  \mu} \right)$ which is assumed to satisfy  the field
equation of $f(T)$. Therefore, for ${{\bar \Gamma}^\mu}_{\nu \rho}$  (which is effected
by LLT) to be identical with  ${{\Gamma}^\mu}_{\nu \rho}$
(which is assumed to satisfies the field equation of $f(T)$)  we must have
\begin{equation}
\left({{\bar \Gamma}^\mu}_{\ \nu \rho}\right)_{Regularized}=\eta^{i j} \left({{ \bar h}_i}^{\ \mu} \right)
\left({{ \bar h}}_{\ j \nu,\;\rho} \right)-
\left({\Lambda_j}^i(x)\right)\left({h_i}^{ \ \mu} \right)\left({\Lambda^j}_k(x)\right)_{,\;\rho}\left({h^k}_{\nu} \right).\end{equation}
 Eq. (23) means that the affine connection is invariant under LLT in the linear case, i.e., $f(T)=T$, which means that the extra degrees of freedom, six ones, are controlled. Also Eq. (23) breaks the  restriction of teleparallelism.
From Eq. (23) we have \[ \Big(\bar{\Gamma}^{\mu}_{\nu \rho}\Big)_{Regularized} \equiv \Gamma^{\mu}_{\nu \rho} \;.\] Therefore,  if   $\Gamma^{\mu}_{\nu \rho} $ satisfies the field equations of $f(T)$ then $\Big(\bar{\Gamma}^{\mu}_{\nu \rho}\Big)_{Regularized}$ need not to be a solution to the field equations of $f(T)$   given by Eq. (8). The main reason for this is the second term in Eq. (8), i.e., $h^{-1}{h^i}_\mu\partial_\rho\left(h{h_i}^\alpha
{S_\alpha}^{\rho \nu}\right)$. This term depends on the  choice of the tetrad field. Using Eq~ (23), the torsion, the contortion  and  ${{\bar S}_\alpha}^{\mu \nu}$  tensors  have the form
\begin{eqnarray} && \left({{\bar T}^\mu}_{\ \nu \rho}\right)_{Regularized}={{\bar T}^\mu}_{\nu \rho}+\left({h_i}^{ \ \mu} \right)
\left({\Lambda^i}_j(x)\right)\left\{\left({\Lambda^j}_k(x)\right)_{,\;\nu}\left({h^k}_{\rho} \right)
-\left({\Lambda^j}_k(x)\right)_{,\;\rho}\left({h^k}_{\nu} \right)\right\},\nonumber\\
&&  \left({{\bar K}^{\ \mu \nu}}_\alpha\right)_{Regularized} =
-\frac{1}{2}\left[\left({{\bar T}^{\ \mu \nu}}_\alpha\right)_{Regularized}-\left({{\bar T}^{\ \nu
\mu}}_\alpha\right)_{Regularized}-\left({{\bar T}_\alpha}^{\ \mu \nu}\right)_{Regularized}\right], \nonumber\\
&&   \left({{\bar S}_\alpha}^{\ \mu \nu}\right)_{Regularized}
= \frac{1}{2}\left[\left({{\bar K}^{\ \mu
\nu}}_\alpha\right)_{Regularized}+\delta^\mu_\alpha\left({{\bar T}^{\ \beta
\nu}}_\beta\right)_{Regularized}-\delta^\nu_\alpha\left({{\bar T}^{\ \beta \mu}}_\beta\right)_{Regularized}\right].\nonumber\\
&&  \end{eqnarray}
Equation (24) shows that  the torsion tensor (and all tensors constructed from it) is invariant
 under LLT. Using equation (24) in the field equations
 of $f(T)$, one can easily see that the first, third and fourth terms of the field
equations (8) will be invariant under LLT, but the second term, $\partial_\rho\left(h{{\bar h}_a}^{\ \alpha}
{{\bar S}_\alpha}^{\ \rho \nu}\right)$, which depends one the derivative  must take the following form
\begin{equation} \left(\partial_\rho\left[h{{\bar h}_a}^{\ \alpha}
{{\bar S}_\alpha}^{\ \rho \nu}\right]\right)_{Regularized}= \partial_\rho\left(h{{\bar h}_a}^{\ \alpha}
{{\bar S}_\alpha}^{\ \rho \nu}\right)-h\left({\Lambda_a}^b(x)\right)_{,\;\rho}\left({{\bar h}_b}^{ \ \alpha} \right) {{\bar S}_\alpha}^{\ \rho \nu}.\end{equation}
Using equations (24) and (25) the invariance field equations of $f(T)$ gravitational theory under LLT take  the form:
\begin{eqnarray}
&& \left({{\bar S}_\mu}^{\ \rho \nu}\right)_{Regularized} \left({\bar T}_{\ ,\rho}\right)_{Regularized} \
f(\bar T)_{{\bar T}{\bar T}}+\Biggl[h^{-1}{{\bar h}^a}_{\ \mu}\left(\partial_\rho\left[h{{\bar h}_a}^{\ \alpha}
{{\bar S}_\alpha}^{\ \rho \nu}\right]\right)_{Regularized}\nonumber\\
&& -\left({{\bar T}^\alpha}_{\ \lambda \mu}\right)_{Regularized}\left({\bar {S}_\alpha}^{\ \nu
\lambda}\right)_{Regularized}\Biggr]f(\bar T)_{\bar T}+\frac{1}{4}{\delta^\nu}_\mu f(\bar T)=4\pi {{\cal
T}^\nu}_\mu,\end{eqnarray}
where $\left({\bar T}\right)_{Regularized}=\left({{\bar T}^\alpha}_{\ \mu \nu}{{\bar S}_\alpha}^{\ \mu \nu}\right)_{Regularized}$.

 Let us check if Eq.~(26) when applied to the tetrad field (9) will indeed
remove the effect of the inertia which appears in the LLT (13).
 Calculating the necessary components of the modified field equations (26) we get a vanishing quantity of the left hand side. This means that the tetrad field (9) is a solution to the
  $f(T)$ field equations (26) which is invariant under LLT\footnote{The details calculations of the non-vanishing components of the necessary quantities of the modified field equations (26) are given in Appendix A.}.
\newsection{Discussion and conclusion}
In this paper we have addressed the problem of the invariance of the field equations of $f(T)$
gravitational theory  under LLT. We first used  a ``general  tetrad field'' which contained
five unknown functions in $r$. This tetrad field has been studied \cite{Nas} and a special solution has been obtained. This solution is
characterized by its scalar torsion is vanishes.

 We rewrite this tetrad field, ``general  tetrad field'', into two matrices. The first matrix represent a tetrad
fields contains four unknown functions in $r$. This tetrad field has been studied before in \cite{Ngrg} and has been
shown that it represent an exact solution within the framework of $f(T)$ gravitational theories. The
second matrix represent a LLT that satisfies
\begin{equation} \left({\Lambda_i}^j \right) \eta_{j k} \left({\Lambda^k}_m \right)=\eta_{im},\end{equation}
 and contains an arbitrary $H(r)$.

 We have applied the
field equations of $f(T)$ {\it which are not invariant under LLT} to the  general tetrad field.
 We have obtained a set of non-linear differential equations which depend on the $H(r)$. Therefore,
 We have regularized the field equations of $f(T)$ gravitational theory such that it has been  became invariant
 under LLT. Then, we have applied these invariant
field equations to the {\it generalized tetrad field}. We have shown that this {\it general tetrad field} is
 an exact solution to the regularized field equations of $f(T)$ gravitational theory.

  The problem of the non invariance of the field equations of $f(T)$  under LLT is not a trivial task to tackle. The main reason for this is the following:
 We have the following known relation between the Ricci scalar tensor and the scalar torsion \cite{LSB1}
\begin{equation} R=-T-2\nabla^\mu{T^\rho}_{\mu \rho}=T-\frac{2}{h}\partial^\mu(h{T^\rho}_{\mu \rho}).\end{equation}
 Last term in the R.H.S. of Eq. (28) is a total divergence term which has no effect on the field equations of TEGR, i.e., ${\cal
L}({h^a}_\mu)=\int d^4x h\left[\frac{1}{16\pi}T\right]$, from this fact comes the well known name teleparallel equivalent of general relativity. However, this term, divergence term is the main reason that makes the field equations of $f(T)$ non invariance under LLT. Let us explain this for some specific form of $f(T)$. If \begin{eqnarray} f(R)=&& R+R^2\equiv
 \left[-T-2\nabla^\mu{T^\rho}_{\mu \rho}\right]+\left[-T-2\nabla^\mu{T^\rho}_{\mu \rho}\right]^2\nonumber\\
 &&=-T-2\nabla^\mu{T^\rho}_{\mu \rho}+T^2+4\left[\nabla^\mu{T^\rho}_{\mu \rho}\right]^2+4T\nabla^\mu{T^\rho}_{\mu \rho},\end{eqnarray}
last term in the R.H.S. of Eq. (29) is not a total derivative term. Therefore, this  term is  responsible to make $f(R)=R+R^2$ when written in terms of $T$ and $T^2$ is not invariant under LLT in contrast to the linear case, i.e., the form of Eq. (28). Same discussion can be applied to the general form of $f(R)$ and $f(T)$ which shows in general a different between the $f(R)$ and $f(T)$ gravitational theories that makes the field equation of $f(R)$ to be of fourth order and invariant under LLT while $f(T)$ is of second order and not invariant under LLT. Here in this study we tackle the problem of the invariance of the field equations of $f(T)$ under LLT for specific symmetry, spherical symmetry. Although the method achieved in this study can be done for any symmetry however,  we do not have the  general local Lorentz transformation that has axial symmetry or homogenous and isotropic, $\cdots$. This  will be study elsewhere.

{\centerline{\bf Appendix A}}
\begin{center}
{{\bf Calculations of the non-vanishing components of the necessary quantities of the modified field equations (26)}}
\end{center}
The non-vanishing components of $\left({\Lambda_a}^b(x)\right)_{,\rho}$:
\begin{eqnarray} && \left({\Lambda^0}_{0, r}\right)=\frac{H\left({\Lambda^0}_{ 1, r}\right)}{L\sin\theta \cos\phi} =
\frac{H'\left({\Lambda^0}_{ 1 ,\theta}\right)}{L\cos\theta \cos\phi} =-\frac{H'\left({\Lambda^0}_{ 1, \phi}
\right)}{L\sin\theta \sin\phi} =\frac{H\left({\Lambda^0}_{ 2, r}\right)}{L\sin\theta \sin\phi}= -\frac{H'
\left({\Lambda^0}_{ 2, \theta}\right)}{L\cos\theta \sin\phi}\nonumber\\
&& \nonumber\\
&& = \frac{H'\left({\Lambda^0}_{ 2, \phi}\right)}{L\sin\theta \cos\phi} =\frac{H\left({\Lambda^0}_{ 3, r}\right)}{L\cos\theta}  =-\frac{H'\left({\Lambda^0}_{ 3, \theta}
\right)}{L\sin\theta }=-\frac{H\left({\Lambda^1}_{0, r}\right)}{L\sin\theta\cos\phi}=-\frac{H'
\left({\Lambda^1}_{0, \theta}\right)}{L\cos\theta\cos\phi}=
\frac{H'\left({\Lambda^1}_{0, \phi}\right)}{L\sin\theta\sin\phi}\nonumber\\
&& \nonumber\\
&& =-\frac{\left({\Lambda^1}_{1, r}\right)}{\sin^2\theta\cos^2\phi}=-\frac{\left({\Lambda^1}_{2, r}\right)}{\sin^2\theta\cos\phi\sin\phi}=-\frac{H\left({\Lambda^2}_{0, r}
\right)}{L\sin\theta\sin\phi}=-\frac{H'\left({\Lambda^2}_{0, 2}\right)}{L\cos\theta\sin\phi}=-\frac{H'
\left({\Lambda^2}_{0, \phi}\right)}{L\sin\theta\cos\phi}\nonumber\\
&& \nonumber\\
&& =-\frac{\left({\Lambda^2}_{3, r}\right)}{\sin\theta\cos\theta\sin\phi}=-\frac{\left({\Lambda^2}_{1, r}\right)}{\sin^2\theta\cos\phi\sin\phi}=-\frac{\left({\Lambda^3}_{1, r}
\right)}{\sin\theta\cos\theta\cos\phi}=-\frac{\left({\Lambda^1}_{3, r}\right)}{\sin\theta\cos\theta \cos\phi}\nonumber\\
&& \nonumber\\
&&=-
\frac{\left({\Lambda^3}_{2, r}\right)}{\sin\theta\cos\theta \sin\phi}=-\frac{\left({\Lambda^3}_{3, r}\right)}{\cos^2\theta}=-\frac{H\left({\Lambda^3}_{0, r}\right)}{L\cos\theta}=-\frac{H'\left({\Lambda^3}_{0, \theta}
\right)}{L\sin\theta}=\displaystyle\frac{HH'}{L},\nonumber\\
&& \nonumber\\
&& \left({\Lambda^1}_{1, \theta}\right)=-\cot\theta \cot\phi \left({\Lambda^1}_{1, \phi}\right)=-\cot\theta \cot\phi
 \left({\Lambda^2}_{2, \phi}\right)=\cot\phi \left({\Lambda^1}_{2, \theta}\right)=\cot\phi \left({\Lambda^2}_{1, \theta}\right) \nonumber\\
&& \nonumber\\
&&=
2\sec 2\phi \cos^2\phi \cot\theta\left({\Lambda^1}_{2, \phi}\right)=
2\sec 2\phi \cos^2\phi \cot\theta\left({\Lambda^2}_{1, \phi}\right)=\tan2\theta \cos\phi
\left({\Lambda^1}_{3, \theta}\right)\nonumber\\
&& \nonumber\\
&&  =-2 \cos^2\phi \csc \phi \left({\Lambda^1}_{3, \phi}\right) =
2\cot^2\phi \cot\theta\left({\Lambda^2}_{2, r}\right) =
\cot^2\phi \left({\Lambda^2}_{2, \theta}\right)    =\tan2\theta \cos^2\phi \csc\phi
\left({\Lambda^2}_{3, \theta}\right)\nonumber\\
&&\nonumber\\
&&= 2\cos\phi  \left({\Lambda^2}_{3, \phi}\right)=\tan2\theta \cos\phi
\left({\Lambda^3}_{1, \theta}\right)=-2\cos^2\phi \csc\phi \left({\Lambda^3}_{1, \phi}\right)=2\tan 2\theta \cos^2\phi \csc\phi \left({\Lambda^3}_{2, \theta}
\right)\nonumber\\
&& \nonumber\\
&& =2\cos\phi  \left({\Lambda^3}_{2, \phi}\right)=-\cos^2\phi \left({\Lambda^3}_{3, \theta}\right)=-\sin2\theta\cos^2\phi(L-1).\end{eqnarray}
The non-vanishing components of the non-symmetric affine connection $\left({{\bar \Gamma}^\mu}_{\nu \rho}\right)_{Regularized}$:
\begin{eqnarray}
&& \left({{\bar \Gamma}^0}_{\ 0 1}\right)_{Regularized}=-\displaystyle\frac{(r-2M)}{r}{{\bar \Gamma}^0}_{\ 1 1}=
\displaystyle\frac{1}{r^2}\left({{\bar \Gamma}^0}_{\ 2 2}\right)_{Regularized}=\displaystyle\frac{1}{r^2\sin^2\theta}
\left({{\bar \Gamma}^0}_{\ 3 3}\right)_{Regularized}=\displaystyle\frac{M}{r(r-2M)},  \nonumber\\
&& \nonumber\\
&& \left({{\bar \Gamma}^1}_{\ 0 1}\right)_{Regularized}=-\left({{\bar \Gamma}^2}_{\ 0 2}\right)_{Regularized}=-\left({{\bar \Gamma}^3}_{\ 0 3}\right)_{Regularized}=-
\left({{\bar \Gamma}^3}_{\ 0 3}\right)_{Regularized}\nonumber\\
&&=\displaystyle\frac{(r-2M)\left({{\bar \Gamma}^1}_{\ 1 1}\right)_{Regularized}}{r}=
\displaystyle\frac{\left({{\bar \Gamma}^1}_{\ 2 2}\right)_{Regularized}}{r^2(r-M)}=\displaystyle\frac{M\left({{\bar \Gamma}^1}_{\ 3 3}\right)_{Regularized}}{r^2\sin^2\theta(r-M)}=-
\displaystyle\frac{M}{r^2}, \nonumber\\
&& \nonumber\\
&& \left({{\bar \Gamma}^2}_{\ 1 2}\right)_{Regularized}=\left({{\bar \Gamma}^3}_{\ 1 3}\right)_{Regularized}=\displaystyle\frac{r-M}{r(r-2M)},
\qquad \left({{\bar \Gamma}^2}_{\ 2 1}\right)_{Regularized}=\left({{\bar \Gamma}^3}_{\ 3 1}\right)_{Regularized}=\displaystyle\frac{1}{r},
\nonumber\\
 && \left({{\bar \Gamma}^2}_{\ 3 3}\right)_{Regularized}=-\sin\theta \cos\theta, \qquad \left({{\bar \Gamma}^3}_{\ 2 3}\right)_{Regularized}=\left({{\bar \Gamma}^3}_{\ 3 2}\right)_{Regularized}=\cot\theta,\nonumber\\
&& \nonumber\\
&& \end{eqnarray}
The non-vanishing components of the torsion $\left({\bar T^\mu}_{\ \nu \rho}\right)_{Regularized}$:
\begin{eqnarray} &&\left({\bar T^1}_{\ 1 0}\right)_{Regularized}=-\left({\bar T^1}_{\ 0 1}\right)_{Regularized}=-\left({\bar T^2}_{\ 2 0}\right)_{Regularized}
=\left({\bar T^2}_{\ 0 2}\right)_{Regularized}=-\left({\bar T^3}_{\ 3 0}\right)_{Regularized}\nonumber\\
&& =\left({\bar T^3}_{\ 0 3}\right)_{Regularized}=-
\displaystyle\frac{M}{r^2}, \nonumber\\
&& \nonumber\\
&& \left({\bar T^0}_{\ 0 1}\right)_{Regularized}=-\left({\bar T^0}_{\ 1 0}\right)_{Regularized}=\left({\bar T^2}_{\ 1 2}\right)_{Regularized}
=-\left({\bar T^2}_{\ 2 1}\right)_{Regularized}=\left({\bar T^3}_{\ 1 3}\right)_{Regularized}\nonumber\\
&& =-\left({\bar T^3}_{\ 3 1}\right)_{Regularized}
=-\displaystyle\frac{M}{r(r-2M)}. \end{eqnarray}
The non-vanishing components of ${{\bar S}_\alpha}^{\ \rho \nu}$:
\begin{eqnarray}
 && \left({{\bar S}_0}^{\ 0 1}\right)_{Regularized}=-\left({{\bar S}_0}^{\ 1 0}\right)_{Regularized}=-\displaystyle\frac{M}{r^2},\nonumber\\
  &&  \left({{\bar S}_1}^{\ 1 0}\right)_{Regularized}=-\left({{\bar S}_1}^{1 0}\right)_{Regularized}=-\displaystyle\frac{M}{r(r-2M)}. \end{eqnarray}
 The non-vanishing components of $\partial_\lambda\left[h{{\bar h}_a}^{\ \alpha}
{{\bar S}_\alpha}^{\ \rho \nu}\right]={{{N}_a}^{\rho \nu}}_{,\;\lambda}$:
\begin{eqnarray} && {{{N}_0}^{0 1}}_{,\;\theta}=-{{{N}_0}^{1 0}}_{,\;\theta}=M(L-H)\cos\theta, \nonumber\\
&& \nonumber\\
&& {{{N}_1}^{1 0}}_{,\;\theta}=-{{{N}_1}^{0 1}}_{,\;\theta}=\cot\phi{{{N}_2}^{1 0}}_{,\theta}
=-\cot\phi{{{N}_2}^{0 1}}_{,\;\theta}=M(L+1-H)\sin\theta \cos\theta\cos\phi, \nonumber\\
&& \nonumber\\
&&
{{{N}_3}^{1 0}}_{,\;\theta}=-{{{N}_3}^{0 1}}_{,\;\theta}=M(L-\tan^2\theta-H)\cos^2\theta, \nonumber\\
&& \nonumber\\
&&
{{{N}_1}^{1 0}}_{,\;\phi}=-{{{N}_1}^{0 1}}_{,\;\phi}=-\tan\phi{{{N}_2}^{1 0}}_{,\phi}
=\tan\phi{{{N}_2}^{0 1}}_{,\;\phi}=-M\sin^2\theta\sin\phi. \end{eqnarray}  Using Eq~(32) in Eq. (25) we get
a vanishing components of $\left(\partial_\rho\left[h{{\bar h}_a}^{\ \alpha}
{{\bar S}_\alpha}^{\ \rho \nu}\right]\right)_{Regularized}$.

\end{document}